\documentclass[twocolumn,aps,prl,superscriptaddress,nofootinbib,showpacs,floatfix,longbibliography]{revtex4-1}
\usepackage{url}
\usepackage{cancel}
\usepackage[colorlinks,linkcolor=blue,citecolor=blue,filecolor=black,urlcolor=blue]{hyperref}
\usepackage{epsfig,graphics}
\usepackage{graphicx}
\usepackage{dcolumn}
\usepackage{bm}
\usepackage[usenames]{color}
\usepackage{amssymb}
\usepackage{amsmath}
\usepackage{multirow}
\usepackage{float}
\usepackage{harpoon}
\usepackage{MnSymbol}
\usepackage{appendix}
\usepackage{color}
\usepackage{hyperref}
\usepackage{cleveref}

\newcommand{\sqrtsnn}{\mbox{$\sqrt{s_{\mathrm{NN}}}$}}
\newcommand{\pT} {p_{\mathrm{T}}}

\newcommand{\nqp}{N_{\mathrm{quark}}}
\newcommand{\pnbd}{p_{\mathrm{nbd}}}
\newcommand{\Nch} {N_{\mathrm{ch}}}
\newcommand{\nch}{N_{\mathrm{ch}}}
\newcommand{\npart}{N_{\mathrm{part}}}

\usepackage{CJK}
\hfuzz=\maxdimen
\begin{document}
\title{Scaling approach to nuclear structure in high-energy heavy-ion collisions}
\newcommand{\sbu}{Department of Chemistry, Stony Brook University, Stony Brook, NY 11794, USA}
\newcommand{\bnl}{Physics Department, Brookhaven National Laboratory, Upton, NY 11976, USA}
\author{Jiangyong Jia}\email[Correspond to\ ]{jiangyong.jia@stonybrook.edu}\affiliation{\sbu}\affiliation{\bnl}
\author{Chunjian Zhang}\email[Correspond to\ ]{chun-jian.zhang@stonybrook.edu}\affiliation{\sbu}
\begin{abstract}
In high-energy heavy-ion collisions, the initial condition of the produced quark-gluon plasma (QGP) and its evolution are sensitive to collective nuclear structure parameters describing the shape and radial profiles of the nuclei. We find a general scaling relation between these parameters and many experimental observables such as elliptic flow, triangular flow, and particle multiplicity distribution. In particular, the ratios of observables between two isobar systems depend only on the differences of these parameters, but not on the details of the final state interactions, hence offering a new way to constrain the QGP initial condition. Using this scaling relation, we show how the structure parameters of $^{96}_{44}$Ru and $^{96}_{40}$Zr conspire to produce the rich centrality dependences of these ratios, as measured by the STAR Collaboration. Our scaling approach demonstrates that isobar collisions are a precision tool to probe the initial condition of heavy-ion collisions, as well as the collective nuclear structures, including the neutron skin, of the atomic nuclei across energy scales. 
\end{abstract}
\maketitle

One main challenge in nuclear physics is to map out the shape and radial structure of the atomic nuclei and understand how they emerge from the interactions among the constituent nucleons~\cite{Nakatsukasa:2016nyc,Nazarewicz:2016gyu}. Varying the number of nucleons along isotopic/isotonic chain often induces rich and non-monotonic changes in the nuclear structure properties. In certain regions of nuclear chart, for example, even adding or subtracting a few nucleons can induce significant deformations and/or changes in the nuclear radius or neutron skin~\cite{Moller:2015fba,Cao:2020rgr,Angeli:2013epw,Centelles:2008vu}. Experimental information on nuclear structure is primarily obtained by spectroscopic or scattering measurements at low energies. But studies show that nuclear structure can be probed in high-energy nuclear collisions at the relativistic heavy ion collider (RHIC) and the large hadron collider (LHC)~\cite{Tanihata:1985psr,Rosenhauer:1986tn,Li:1999bea,Heinz:2004ir,Filip:2009zz,Shou:2014eya,Goldschmidt:2015kpa,Giacalone:2017dud,Giacalone:2019pca,Giacalone:2020awm,Giacalone:2021uhj,Giacalone:2021udy,Jia:2021wbq,Jia:2021tzt,Bally:2021qys}, and experimental evidences have been observed~\cite{Adamczyk:2015obl,Acharya:2018ihu,Sirunyan:2019wqp,Aad:2019xmh,jjia}. 

\begin{figure}[!h]
\includegraphics[width=0.9\linewidth]{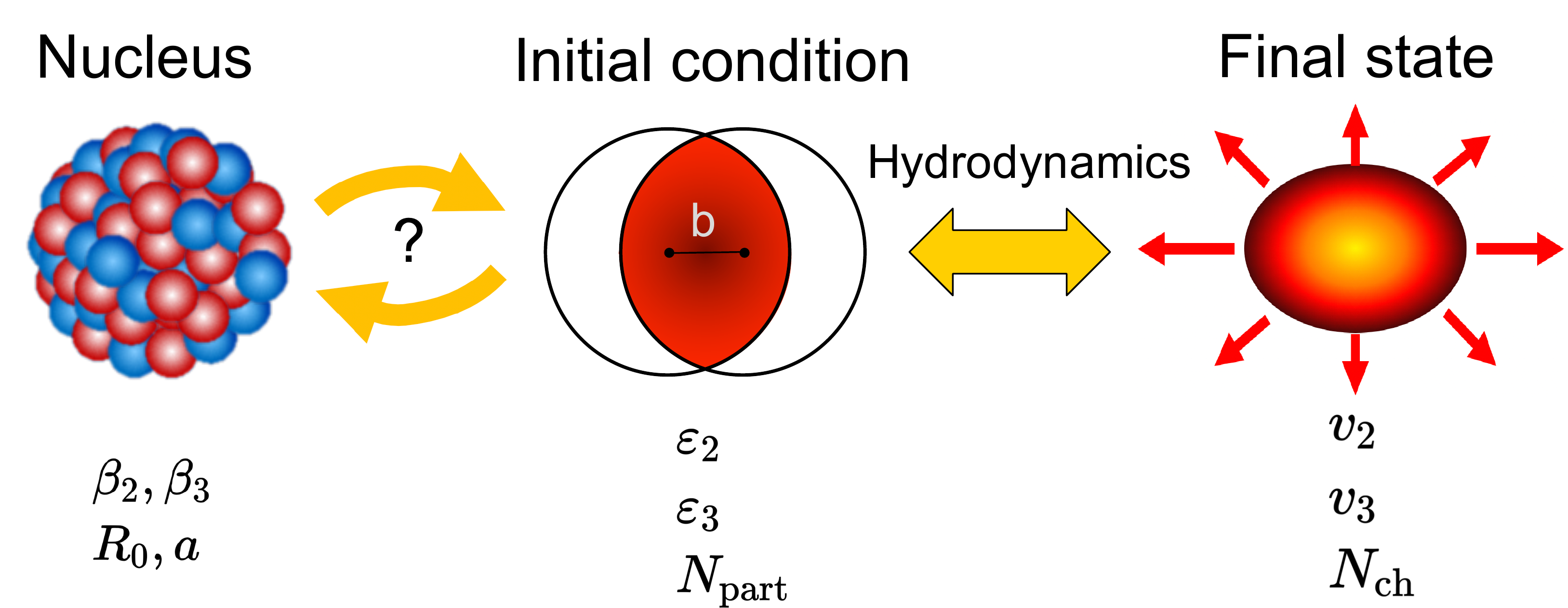}
\vspace*{-.3cm}
\caption{\label{fig:0} Connection between collective nuclear structure (left), the initial condition (middle) and final state (right) of high-energy heavy-ion collisions, together with parameters describing the geometrical aspects for each phase (see text). Due to the extremely short nuclear crossing time $2R_0/\gamma\lesssim0.1$~fm/$c$, the initial condition is well separated from the hydrodynamical evolution in the final state. The geometry of the initial condition depends on the impact parameter $b$ and structure parameters.}
\end{figure}

The connection between nuclear structure and high-energy heavy-ion collisions is illustrated in Fig.~\ref{fig:0}. These collisions deposit a large amount of energy in the overlap region in the middle panel, forming a hot and dense quark-gluon plasma (QGP)~\cite{Busza:2018rrf}. Driven by large pressure gradients, the QGP undergoes a hydrodynamical expansion, converting the initial spatial anisotropies into momentum anisotropies of particles in the final state in the right panel. Observables describing the collective features of the particles in the final state, such as elliptic flow $v_2$, triangular flow $v_3$ and charged particle multiplicity $\nch$ are closely related to geometric features of the initial condition, ellipticity $\varepsilon_2$, triangularity $\varepsilon_3$ and number of participating nucleons $\npart$, respectively. In fact, at energies reached at RHIC and the LHC, $\sqrtsnn\geq 100$ GeV, these quantities are linearly-related $v_n\propto \varepsilon_n$ and $\nch\propto \npart$~\cite{Teaney:2012ke,Niemi:2015qia}. On the other hand, the shape and size of the initial condition are affected by the nucleon distribution in the colliding nuclei in the left panel, often described by a deformed Woods-Saxon (WS) density,
\begin{align}\label{eq:f1}
\rho(r,\theta,\phi)&\propto\frac{1}{1+e^{[r-R_0\left(1+\beta_2 Y_2^0(\theta,\phi) +\beta_3 Y_3^0(\theta,\phi)\right)]/a}},
\end{align}
containing four structure parameters: quadrupole deformation $\beta_2$ and octupole deformation $\beta_3$, half-density radius $R_0$, and surface diffuseness $a$~\cite{Horiuchi:2021dku}. The deformation $\beta_2$ ($\beta_3$) enhances the $\varepsilon_2$ ($\varepsilon_3$) in the initial condition~\cite{Heinz:2004ir,Carzon:2020xwp,Jia:2021tzt}. A change in $a$ influences $\varepsilon_n$ and charge particle multiplicity distribution $p(\npart)$~\cite{Shou:2014eya,Li:2018oec}. Both $a$ and $R_0$ were shown to have significant impact on the initial overlap area~\cite{Zhang:2022fou,Xu:2021uar}. In more recent studies, these structure parameters are found to have much larger impact on multi-point correlators in both the initial and final state~\cite{Zhao:2022uhl,Jia:2022qrq,Jia:2022qgl}. Understanding the role of nuclear structure can improve modeling of the initial condition, which currently limits the extraction of the transport properties of the QGP~\cite{Bernhard:2016tnd,Everett:2020xug,Nijs:2020ors}. 

Due to the dominant role of the impact parameter, earlier studies focused on the most central collisions where the impact of nuclear structure can be easily identified. It is realized recently that the nuclear structure impact can be cleanly isolated over the full centrality range by comparing two isobaric collision systems~\cite{Giacalone:2021uhj,Giacalone:2021udy}.  Since isobar nuclei have the same mass number but different structures, deviation from unity of the ratio of any observable must originate from differences in the structure of the colliding nuclei, which impact the initial state of QGP and its final state evolution. Collisions of one such pair of isobar systems, $^{96}$Ru+$^{96}$Ru and $^{96}$Zr+$^{96}$Zr, have been performed at RHIC. Ratios of many observables are found to show significant and centrality-dependent departures from unity~\cite{STAR:2021mii}.  The goal of this Letter is to explore the scaling behavior of these ratios with respect to the WS parameters in Eq.~\eqref{eq:f1}. 

We illustrate this point using three heavy-ion observables, the $v_2(\nch)$, $v_3(\nch)$ and $p(\nch)$, although the same idea applies to many other single-particle or two-particle observables. For small deformations and small variations of $R_0$ and $a$ from their default reference values, the observable $\mathcal{O}$ has the following leading-order form, 
\begin{align}\label{eq:f1b}
\mathcal{O} \approx b_0+b_1\beta_2^2+b_2\beta_3^2+b_3 (R_0-R_{0,\mathrm{ref}})+b_4(a-a_{\mathrm{ref}})\;,
\end{align}
where $b_0$ is the value for spherical nuclei at some reference radius and diffuseness, and $b_1$--$b_4$ are centrality-dependent response coefficients that encode the final-state dynamics.~\footnote{Note that the leading-order contribution from deformation appears as $\beta_n^2$ instead of $\beta_n$ because these observables do not depend on the sign of $\beta_n$~\cite{Jia:2021qyu,Jia:2021tzt} For higher-order correlators, such as skewness of $\pT$ fluctuation and $v_n^2-\pT$ correlation, the leading order term scales with $\beta_n^3$~\cite{Jia:2021qyu}.}. Most dependence on mass number is carried by $b_0$, while $b_1$--$b_4$ are expected to be weak functions of mass number. The ratio of $\mathcal{O}$ between $^{96}$Ru+$^{96}$Ru and $^{96}$Zr+$^{96}$Zr then has a simple scaling relation
\begin{align}\label{eq:f2}
R_{\mathcal{O}}\equiv\frac{\mathcal{O}_{\mathrm{\mathrm{Ru}}}}{\mathcal{O}_{\mathrm{\mathrm{Zr}}}} \approx 1+ c_1 \Delta\beta_2^2 +c_2 \Delta\beta_3^2 + c_3\Delta R_0 +c_4\Delta a\;,
\end{align}
where $\Delta\beta_n^2= \beta_{n,\mathrm{\mathrm{Ru}}}^2-\beta_{n,\mathrm{\mathrm{Zr}}}^2$, $\Delta R_0 =R_{0,\mathrm{\mathrm{Ru}}}-R_{0,\mathrm{\mathrm{Zr}}}$, $\Delta a =a_{\mathrm{\mathrm{Ru}}}-a_{\mathrm{\mathrm{Zr}}}$ and $c_n=b_n/b_0$. Two important insights can be drawn if Eq.~\eqref{eq:f2} holds: 1) these ratios can only probe the difference in the WS parameters between the isobar nuclei, 2) the contributions are independent of each other among the WS parameters.

To verify this scaling relation, we simulate the dynamics of the QGP using the multi-phase transport model (AMPT)~\cite{Lin:2004en}. The AMPT model describes collective flow data at RHIC and the LHC~\cite{Xu:2011fi,STAR:2021twy} and was used to study the $\beta_n$ dependence of $v_n$~\cite{Giacalone:2021udy,Zhang:2021kxj}. We use AMPT v2.26t5 in string-melting mode at $\sqrtsnn=200$~GeV with a partonic cross section of 3.0~$m$b~\cite{Ma:2014pva,Bzdak:2014dia}. We simulate generic isobar $^{96}$X+$^{96}$X collisions covering a wide range of $\beta_2$, $\beta_3$, $R_0$ and $a$, including the default values assumed for $^{96}$Ru and $^{96}$Zr listed in Table~\ref{tab:1}. Following Ref.~\cite{Xu:2021vpn}, the default values are taken from Ref.~\cite{1995ADNDT} for $R_0$ or deduced from neutron skin data~\cite{Trzcinska:2001sy} for $a$.  The default values of $\beta_2$ and $\beta_3$ are taken from Ref.~\cite{Zhang:2021kxj}. The $v_n$ are calculated using two-particle correlation method with hadrons of $0.2<\pT<2$ GeV and $|\eta|<2$~\cite{ATLAS:2012at}. The ratios are calculated as a function of $\nch$ instead of centrality, because the ratios calculated at the same $\nch$ have a good cancellation of non-flow contributions~\cite{Jia:2022iji} and the final state effects~\cite{Zhang:2022fou}.

\begin{table}[!h]
\centering
\begin{tabular}{|c|cccc|}\hline 
\text {Species} & $\beta_{2}$ & $\beta_{3}$ & \;$a$\; &\; $R_0$\; \\\hline 
$^{96}$Ru & 0.162 & 0 & 0.46 fm & 5.09 fm \\\hline 
$^{96}$Zr & 0.06 & 0.20 & 0.52 fm & 5.02 fm \\\hline\hline
\multirow{2}{*}{difference} & \;\;$\Delta\beta_{2}^2$\;\; & \;\;$\Delta\beta_{3}^2$\;\; & \;\;\;$\Delta a$ \;\;\; & \;\;\;$\Delta R_0$\;\;\; \\\cline{2-5} 
 & 0.0226 & -0.04 & -0.06 fm & 0.07 fm \\\hline
\end{tabular}
\caption{\label{tab:1} Collective nuclear structure parameters for $^{96}$Ru and $^{96}$Zr and the differences.} 
\end{table}

\begin{figure*}
\includegraphics[width=1\linewidth]{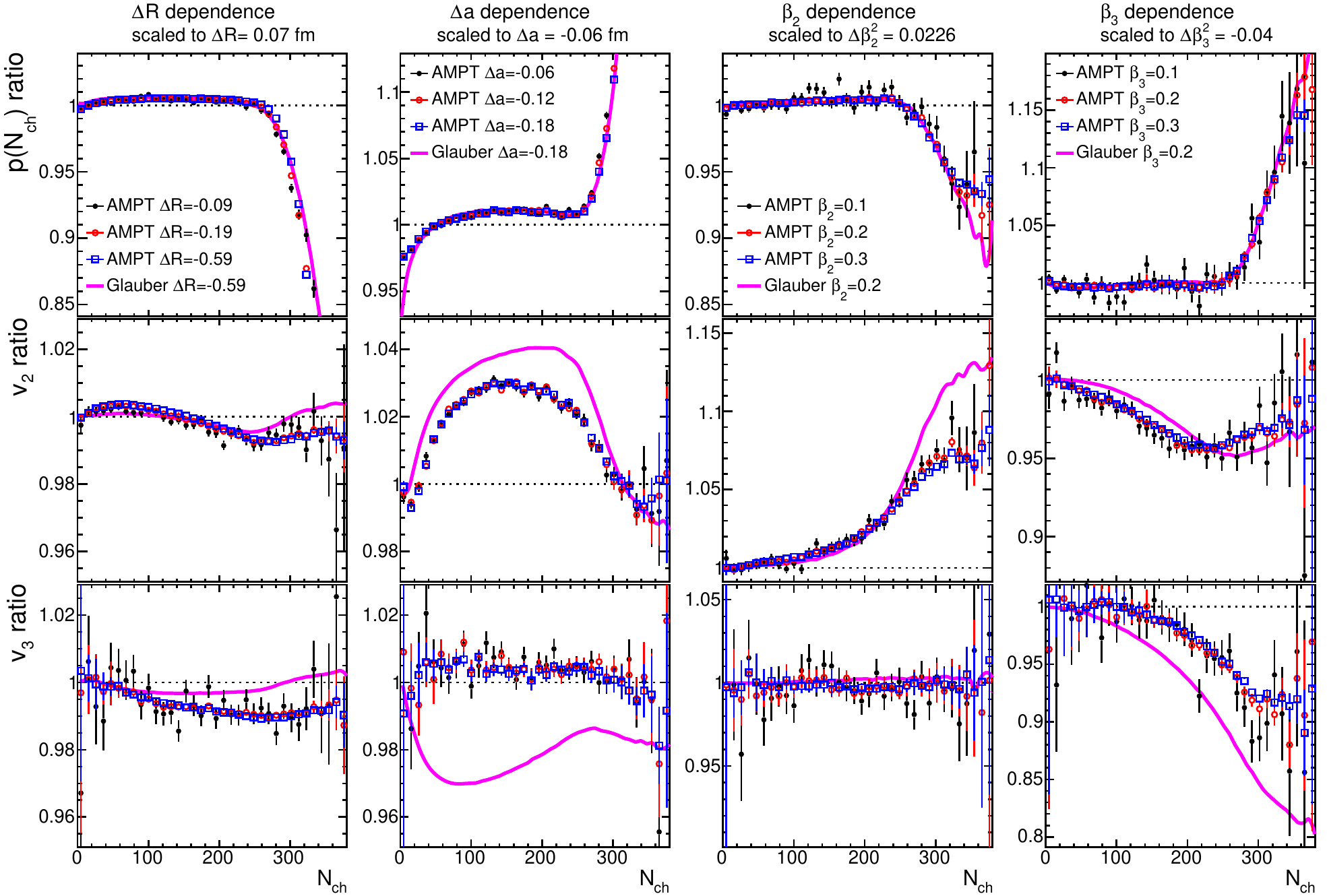}
\vspace*{-.4cm}
\caption{\label{fig:1} The four terms of Eq.~\eqref{eq:f2} associated with $R_0$ (left column), $a$ ($2^{\mathrm{nd}}$ column), $\beta_2$ ($3^{\mathrm{rd}}$  column) and $\beta_3$ (right column) from the AMPT model for ratios of $p(\nch)$ (first row), $v_2$ (middle row) and $v_3$ (bottom row). Distribution in each panel is determined for several values of parameters and scaled to the same default value. They are compared with those obtained for quark Glauber model (solid lines).}
\end{figure*}

To explore the parametric dependence of the hydrodynamic response, the parameters are varied one at a time. The $\beta_2$ is changed from 0 to 0.1, 0.15 and 0.2; the $\beta_3$ is changed from 0 to 0.1, 0.2 and 0.25; the $a$ is varied from 0.52 fm to 0.46 fm, 0.40 fm and 0.34 fm; the $R_0$ is varied from 5.09 fm to 5.02 fm, 4.8 fm and 4.5 fm. An independent sample is generated for each case and the $v_2$, $v_3$ and $p(\nch)$ are calculated. The change in the ratios from unity, $R_{\mathcal{O}}-1$, are scaled according to the actual differences between Ru and Zr listed in Table~\ref{tab:1}. The results for all twelve cases (four parameters times three observables) as a function of $\nch$ are summarized in Fig.~\ref{fig:1}.
\begin{figure*}
\includegraphics[width=1\linewidth]{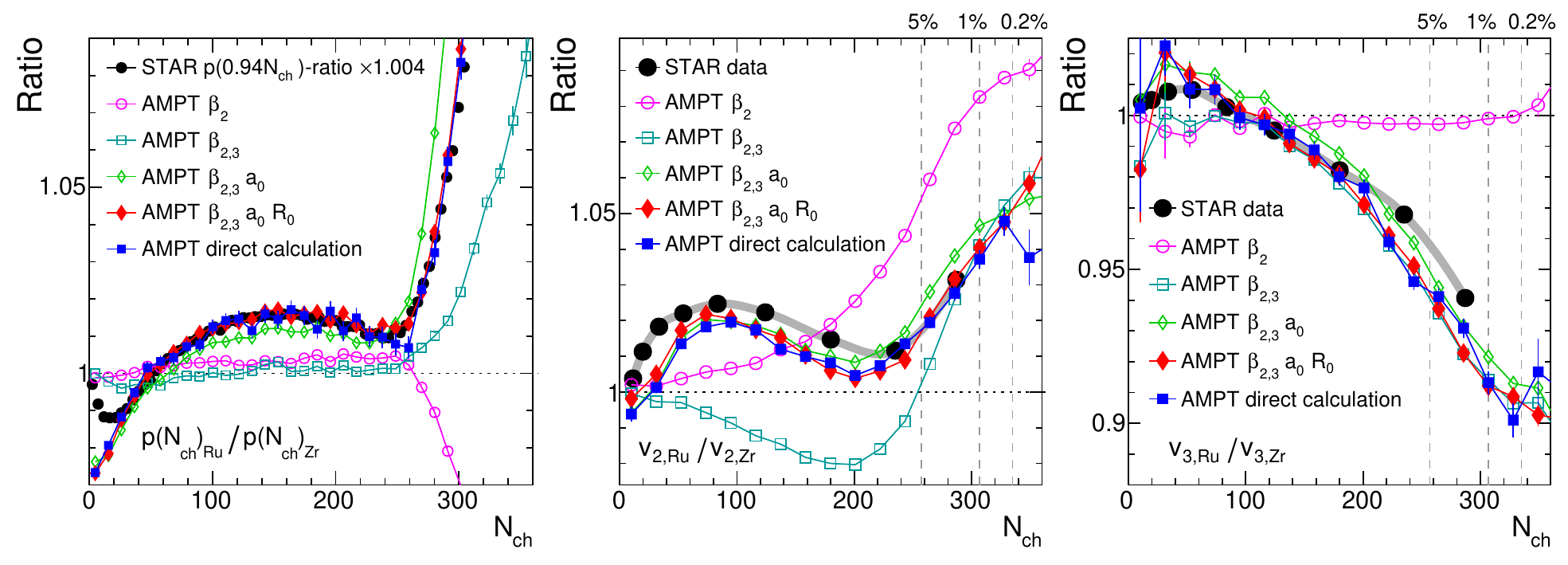}
\vspace*{-.4cm}
\caption{\label{fig:2} The ratios of $p(\nch)$ (left panel), $v_2$ (middle panel), and $v_3$  (right panel) from AMPT using the default nuclear structure parameters in Tab.~\ref{tab:1} (labeled ``direct calculation'') or calculated step-by-step from the response coefficients from Fig.~\ref{fig:1}. The $\nch$ and $R_{p(\nch)}$ values from data have been scaled by 0.94 and 1.004, respectively to match the AMPT. Note also that the STAR $v_n$ ratios at the same centrality~\cite{STAR:2021mii} has been corrected to reflect the ratio at the same $\nch$~\cite{Jia:2022iji}.}
\end{figure*}

One striking feature is the nearly perfect scaling of $R_{\mathcal{O}}$ over the wide range of parameter values studied. The shapes of these dependences reflect directly the response coefficients $c_n(\nch)$ for each observable. The statistical uncertainties of $c_n$ decrease for larger variations of the WS parameters, implying that the $c_n$ can be determined more precisely by using a larger change of each parameter. This has the benefit of significantly reducing the number of events required in the hydrodynamic model simulation to achieve the desired precision, ideally suitable for the multi-system Bayesian global analyses of heavy-ion collisions~\cite{Nijs:2020ors,JETSCAPE:2020shq}.

All the WS parameters do not have the same influences on final state observables. In peripheral and mid-central collisions, the ratio $p(\nch)_{\mathrm{Ru}}/p(\nch)_{\mathrm{Zr}}$ is influenced mostly by the $\Delta a$ and $\Delta R_0$. In particular, the characteristic broad peak and non-monotonic behavior of the ratio is a clear signature of the influence of $\Delta a$~\cite{Xu:2021vpn}. In the most central collisions, the ratio is sensitive to all four parameters. The influence of WS parameters on $v_{2,\mathrm{Ru}}/v_{2,\mathrm{Zr}}$ is more rich: 1) in the most central collisions, the ratio is mainly dominated by $\Delta\beta_2^2$ and to a lesser extent by $\Delta\beta_3^2$;  2) in the near-central collisions, the ratio is influenced by a positive contribution from $\Delta\beta_2^2$ and a larger negative contribution from $\Delta\beta_3^2$; 3) in the mid-central and peripheral collisions, the impact of $\Delta a$ is more important; 4) the influence of $\Delta R_0$ is negligible except in central collisions. Lastly, the ratio $v_{3,\mathrm{Ru}}/v_{3,\mathrm{Zr}}$ is mainly influenced by $\Delta\beta_3^2$, although $\Delta a$ and $\Delta R_0$ have opposite up to 1\% contributions over a broad $\nch$ range. 

The scaling relation in Fig.~\eqref{fig:1} allows us to construct directly the ratios of experimental observables for any values of $\Delta \beta_2^2$, $\Delta \beta_3^2$, $\Delta a$ and $\Delta R_0$, without the need to carry out additional simulations. One could also perform a simultaneous fit of several experimental ratios to obtain the optimal values of these parameters within a given model framework and expose its limitations. Figure~\ref{fig:2} shows a step-by-step construction of the prediction in comparison with the STAR data.  Each panel also shows the ratio obtained directly from a separate AMPT simulation of $^{96}$Ru+$^{96}$Ru and $^{96}$Zr+$^{96}$Zr collisions using the default parameters in the Table~\ref{tab:1}. Excellent agreement is obtained between the construction approach and the direct calculation, attesting to the robustness of our proposed method. Also for the first time, we achieved simultaneous description of all three ratios using one set of WS parameters in most centrality ranges.

One natural question is how these isobar ratios are influenced by various final state effects. A recent study from us has demonstrated explicitly that isobar ratios are insensitive to the shear viscosity, hadronization and hadronic transport~\cite{Zhang:2022fou}. Therefore, any model dependence in the isobar ratios must reflect a model dependence in the initial condition, i.e. how the energy is deposited in the overlap region (see Fig.~\ref{fig:0}). One example is the response functions calculated from a quark Glauber model shown in Fig.~\ref{fig:1} (details in Supplemental materials), which has clear differences in several cases from the AMPT model. There are potentially many initial conditions, reflected by the well-known T$_{\mathrm{R}}$ENTo formula for the energy density $e(x,y) \propto\left(T_{\mathrm{A}}^{p}+T_{\mathrm{B}}^{p}\right)^{q/p}$ calculated from the thickness function $T_{\mathrm{A}}$ and $T_{\mathrm{B}}$ of colliding ions, where each $q$ and $p$ value specify a different initial condition~\cite{Moreland:2014oya,Nijs:2022rme}. The coefficients $c_n$ provide a new way to constrain the initial condition by exploiting structure differences between isobars. One has to first calibrate the values of $c_n$ using species whose WS parameters are relatively well known. The calibrated $c_n$ can then be used to 1) narrow the $q$ and $p$ values, which can be subsequently fixed in the Bayesian inference to improve the extraction of the QGP properties, and 2) constrain the nuclear structure parameters for species of interest by directly fitting Eq.~\eqref{eq:f2} to the measured isobar ratios. 

A caveat is in order regarding to the connection between nuclear structure and initial condition. The parameters describing the shape of the nuclei in high energy may not take the same values as those at low-energy. In fact, nuclear structure at small partonic longitudinal momentum fraction (small-$x$), is expected to be modified due to gluon shadowing or saturation effects, described by nPDF or nuclear partonic distribution function. The nPDF appears as additional spatial modulation of the nucleon distribution in the transverse plane, and will modify the values of the parameters in Eq.~\eqref{eq:f1} in a $x$-dependent way. The nPDF effects, as input to the heavy-ion initial condition, is a key topic in $e$+A collisions at future electron-ion collider (EIC) and $p$+A collisions. The isobar collisions provide a new means to access modification of nuclear structure in dense gluon environment in a data-driven approach, for example by comparing isobar ratios between RHIC and the LHC energies or as a function of rapidity.

The scaling approach discussed above can be extended to compare collisions of systems with similar but slightly different mass number $A$, ideally along an isotopic chain. As the $\nch$ distribution scales approximately with $A$, the ratios of experimental observable can be obtained as a function of $\nch/(2A)$ or centrality.  In this case, one has $b_0\rightarrow b_0 (1+\frac{d\ln b_0}{d\ln A} \frac{\Delta A}{A})$, which leads to one additional term, $\frac{d\ln b_0}{d\ln A} \frac{\Delta A}{A}$, in Eq.~\eqref{eq:f2}. The $A$ dependence of $c_n$ is weak and also its contribution to Eq.~\eqref{eq:f2} has a higher-order form, e.g. $\frac{\Delta A}{A} \Delta a$, etc., therefore is ignored. Studies along this line have been done for elliptic flow~\cite{Giacalone:2021udy,Jia:2021tzt}, which show that the $b_0$ for $\epsilon_2$ has the form $b_0\propto 1/A$ in the ultra-central collisions, and that $R_{\epsilon_2}$ receives an additional correction $-\Delta A/A$. This contribution should be quantified for each observable and compared to data from two systems of similar sizes, such as $^{197}$Au+$^{197}$Au and $^{238}$U+$^{238}$U. In conjunction with the scaling relations for the nuclear structure parameters discussed above, they can be a powerful tool in understanding the system size dependence of heavy-ion observables.

The scaling approach also provides a clean way to probe the difference between the root mean square radius of neutrons and protons in heavy nuclei, $\Delta r_{np}= R_{n}-R_{p}$, known as the neutron skin. The $\Delta r_{np}$ is related to the symmetry energy contribution to the equation of state (EOS): a quantity of fundamental importance in nuclear- and astro-physics~\cite{Lattimer:2006xb,Li:2008gp}. From discussion above, the isobar ratios are expected to probe only the difference in the neutron skin. To see this, we first express the mean square radius of nucleon distribution in Eq.~\eqref{eq:f1} by $R^2\approx(\frac{3}{5}R_0^2+\frac{7}{5}\pi^2a^2)/(1+\frac{5}{4\pi}\sum_n\beta_n^2)$~\cite{bohr}. The neutron skin is then expressed in terms of the differences between nucleon distribution and proton distribution:
\small{\begin{align}\label{eq:f4}
\Delta r_{np}\!=\!\frac{R^2-R_p^2}{R(\delta\!+\!1)}\!&\approx\!\frac{3(R_{0}^2\!-\!R_{0,p}^2)\!+\!7\pi^2(a^2\!-\!a_p^2)}{\sqrt{15}R_0\sqrt{1\!+\!\frac{7\pi^2}{3}\!\frac{a^2}{R_0^2}}(1\!+\!\delta\!+\!\frac{5}{8\pi}\!\sum_n\beta_n^2)},
\end{align}}\normalsize
where $\delta = (N-Z)/A$, and $R_{0,p}$ and $a_p$ are the well-measured WS parameters for the proton distribution~\cite{1995ADNDT}. Simple algebraic manipulation shows that $\Delta R_0$ and $\Delta a$ are related to the skin difference~\cite{supp},
\begin{align}\nonumber
\Delta (\Delta r_{np})\approx-&\overline{\Delta r}_{np}(\frac{\Delta \delta}{1\!+\!\bar{\delta}}+\frac{\Delta R_0}{\bar{R}_0}) +\\\label{eq:f5}
&\frac{\Delta Y -\frac{7\pi^2}{6}\frac{\bar{a}^2}{\bar{R}_0^2}\left(\Delta Y + 2\bar{Y}(\frac{\Delta a}{\bar{a}}-\frac{\Delta R_0}{\bar{R}_0})\right)}{\sqrt{15}\bar{R}_0(1+\bar{\delta}+\frac{5}{8\pi}\sum_n\overline{\beta_n^2})}\;,
\end{align}
where $\bar{x}$ represents the average of $x$ between the two systems, and $Y \equiv 3(R_{0}^2\!-\!R_{0,p}^2)\!+\!7\pi^2(a^2\!-\!a_p^2)$. The term associated with $\overline{\Delta r}_{np}$ can be dropped if we ignore change of $\delta$ and $R_0$, which is typically a few percents of $\overline{\Delta r}_{np}$ for isobar systems. The numerator of Eq.~\eqref{eq:f5} is dominated by $\Delta Y= 6(\bar{R}_0\Delta R_0\!-\!\bar{R}_{0p}\Delta R_{0p})\!+\!14\pi^2(\bar{a}\Delta a\!-\!\bar{a}_p \Delta a_p)$, the remaining term is on the order of $\frac{7\pi^2}{6}\frac{\bar{a}^2}{\bar{R}_0^2}\sim 11 (0.5/5)^2=11\%$ of $\Delta Y$. We checked that the Eq.~\eqref{eq:f5} is accurate within 2\% using parameters for $^{96}$Ru and $^{96}$Zr listed in Ref.~\cite{Xu:2021vpn}. 

Knowledge of nucleon distribution gives direct information on the neutron skin, once it is combined with the well-known proton distribution. Eq.~\eqref{eq:f5} shows that isobar data can only constrain the neutron skin difference, which can be constructed from $\Delta R_0$ and $\Delta a$, together with well-measured $\Delta R_{0,p}$ and $\Delta a_p$ for protons. The neutron skin difference is sensitive to both $\Delta R_0$ (skin-type contribution) and $\Delta a$ (halo-type contribution)~\cite{Trzcinska:2001sy,Centelles:2010qh}. Previous studies of neutron skin are done by inputting density functional theory (DFT) calculation of nuclear structure directly to the hydrodynamic modeling of heavy-ion collisions~\cite{Li:2019kkh,Xu:2021uar,Xu:2021vpn}. The neutron skin values are constrained by comparing directly with experimental observables. What we are proposing here is to decouple DFT from modeling of heavy-ion collisions. One first extracts the $\Delta R_{0}$ and $\Delta a$ values consistent with many isobar ratios using the scaling approach, which are then compared with those calculated directly from nucleon distributions from the DFT theory. Eq.~\eqref{eq:f5} provides an easy way to estimate the skin difference, and contributions from skin-type or halo-type. 

In summary, we presented a new approach to constrain the collective nuclear structure parameters in high-energy heavy-ion isobar collisions. We found that the changes in the final state observables $v_2(\nch)$, $v_3(\nch)$ and $p(\nch)$ follow a simple dependence on the variation of these parameters. The coefficients of these variations can be determined precisely in a given model framework, and subsequently used to make predictions of observables at other parameter values. This scaling behavior is particularly useful in analyzing the ratios between isobar systems, such as $^{96}$Ru+$^{96}$Ru and $^{96}$Zr+$^{96}$Zr collisions measured by the STAR experiment~\cite{STAR:2021mii}. We show that the STAR data can constrain directly the nuclear structure differences between $^{96}$Ru and $^{96}$Zr (compatible with the structure values in Tab.~\ref{tab:1}). Since these isobar ratios are also found to be insensitive to the details of interaction in the final state, the isobar collisions serve as a precise tool for accessing both the bulk nuclear structure parameters and the initial condition of heavy-ion collisions. The extracted information on nucleon distribution, together with well-measured charge distribution, can probe the difference in the neutron skin between large isobar systems. However, future measurements of isobar ratios as a function of collision energy and rapidity, are necessary to quantify the modification of nuclear structure at high-energy across energy scales, and establish more firmly the connection between nuclear structure and the initial condition. Our study demonstrates the unique opportunities offered by relativistic collisions of isobars as a tool to perform inter-disciplinary nuclear physics studies, which we hope will be pursued in future by collisions of several isobar pairs in collider facilities.

{\bf Acknowledgements:} We thank Giuliano Giacalone,  Che-Ming Ko, Bao-An Li and Jun Xu for careful reading and valuable comments on the manuscript. This work is supported by DOE DE-FG02-87ER40331.
\section*{Supplemental materials}
To show how Eq.~\eqref{eq:f5} is derived, we note that the ms radii for nucleon, neutron and proton distributions are related by
\begin{align}\label{eq:a1}
R^2 = \frac{1+\delta}{2} R_n^2+  \frac{1-\delta}{2}R_p^2\;.
\end{align}
This leads to $\Delta r_{np} \frac{R_n+R_p}{2} = \frac{R^2-R_p^2}{1+\delta}$, and together with the approximation $R_n+R_p\approx 2R-\delta \Delta r_{np}$, we get
\begin{align}\label{eq:a2}
\Delta r_{np}(1-\frac{\Delta r_{np}\delta}{2R}) = \frac{R^2-R_p^2}{R(1+\delta)}\;.
\end{align}
Ignoring the $\Delta r_{np}\delta/(2R)$ term, which is typically much less than 1\%, we obtain Eq.~(4). To derive Eq.~(5), we first rewrite Eq.~(4) as,
\begin{align}\nonumber
\Delta r_{np}&\sqrt{15}R_0(1+\delta\!+\!\frac{5}{8\pi}\!\sum_n\beta_n^2) \approx \\\label{eq:a3} 
&(3(R_{0}^2\!-\!R_{0,p}^2)\!+\!7\pi^2(a^2\!-\!a_p^2))(1-\frac{7\pi^2}{6}\frac{a^2}{R_0^2})\;,
\end{align}
where we have ignored the high-order term $\mathcal{O}(\frac{a^4}{R_0^4})$ on the right hand side (rhs), which is much less than 1\% for medium and large nuclei. We now consider the difference of two such expressions for two isobars, labelled by ``1'' and ``2'', respectively. We shall use the relations $x_1y_1-x_2y_2 = \Delta x \bar{y}+\Delta y \bar{x}$ and $x_1y_1z_1-x_2y_2z_2 = \Delta x \bar{y}\bar{z}+\Delta y \bar{x}\bar{z}+\Delta z \bar{x}\bar{y}+\frac{1}{4}\Delta x\Delta y\Delta z$, where $\Delta x = x_1-x_2$ and $\bar{x} = (x_1+x_2)/2$ etc. Then, the left hand side (lhs) of Eq.~\eqref{eq:a3} can be written as,
\begin{align}\nonumber
\Delta (\mathrm{lhs})/\sqrt{15}&=\Delta\left[\Delta r_{np}R_0(1+\delta+\frac{5}{8\pi}\sum_n \beta_n^2)\right]\\\nonumber
&\approx \Delta(\Delta r_{np})\bar{R}_0(1+\bar{\delta}+\frac{5}{8\pi}\sum_n\overline{\beta_n^2})\\\label{eq:a4}
&+\overline{\Delta r}_{np}\bar{R}_0(\frac{\Delta R_0}{\bar{R}_0}(1+\bar{\delta})+\Delta \delta+\frac{5}{8\pi}\!\sum_n\!\Delta\beta_n^2)\;.
\end{align}
For the rhs of Eq.~\eqref{eq:a3}, it can be written as,
\begin{align}\nonumber
\Delta (\mathrm{rhs})&=\Delta \left[(3(R_{0}^2\!-\!R_{0,p}^2)\!+\!7\pi^2(a^2\!-\!a_p^2))(1-\frac{7\pi^2}{6}\frac{a^2}{R_0^2})\right]\\\label{eq:a5}
&=\Delta Y -\frac{7\pi^2}{3}\frac{\bar{a}^2}{\bar{R}_0^2}\left(\frac{\Delta Y}{2} + \bar{Y}(\frac{\Delta a}{\bar{a}}-\frac{\Delta R_0}{\bar{R}_0})\right)\;,
\end{align}
with $Y \equiv 3(R_{0}^2\!-\!R_{0,p}^2)\!+\!7\pi^2(a^2\!-\!a_p^2)$. Eq.~(5) is then obtained by combining Eq.~\eqref{eq:a4} and Eq.~\eqref{eq:a5}.

\begin{figure*}[!t]
\includegraphics[width=0.95\linewidth]{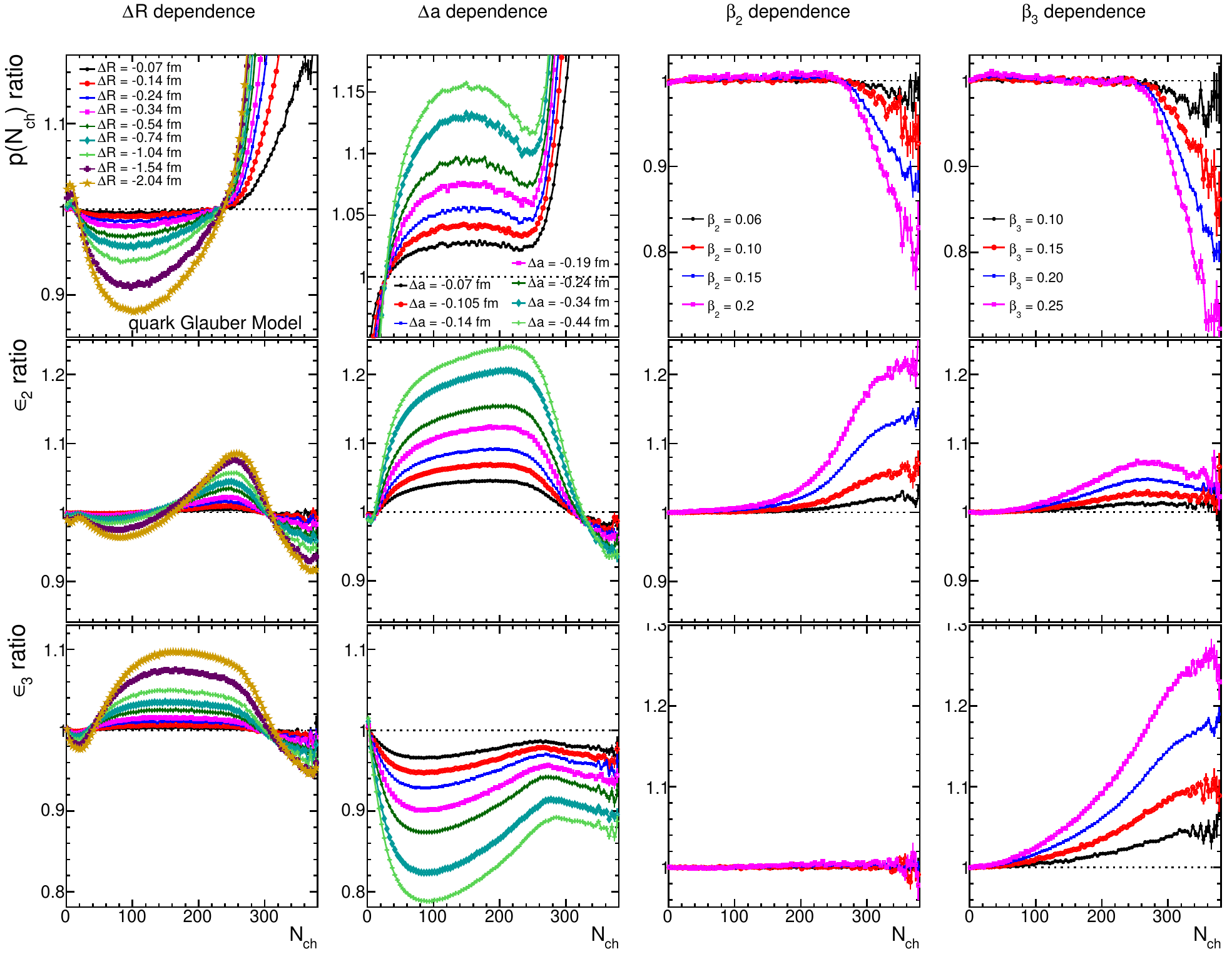}
\caption{\label{fig:4} The four response coefficients of Eq.~(3) for variation of nuclear structure parameters $R_0$ (left column), $a$ (2nd column), $\beta_2$ (third column) and $\beta_3$ (right column) in the quark Glauber model for ratios of $p(\Nch)$ (first row), $\varepsilon_2$ (middle row) and $\varepsilon_3$ (bottom row).  Each coefficient in each panel is determined for several values of the parameters and scaled to the same default value.}
\end{figure*}

\begin{figure*}[!h]
\includegraphics[width=0.95\linewidth]{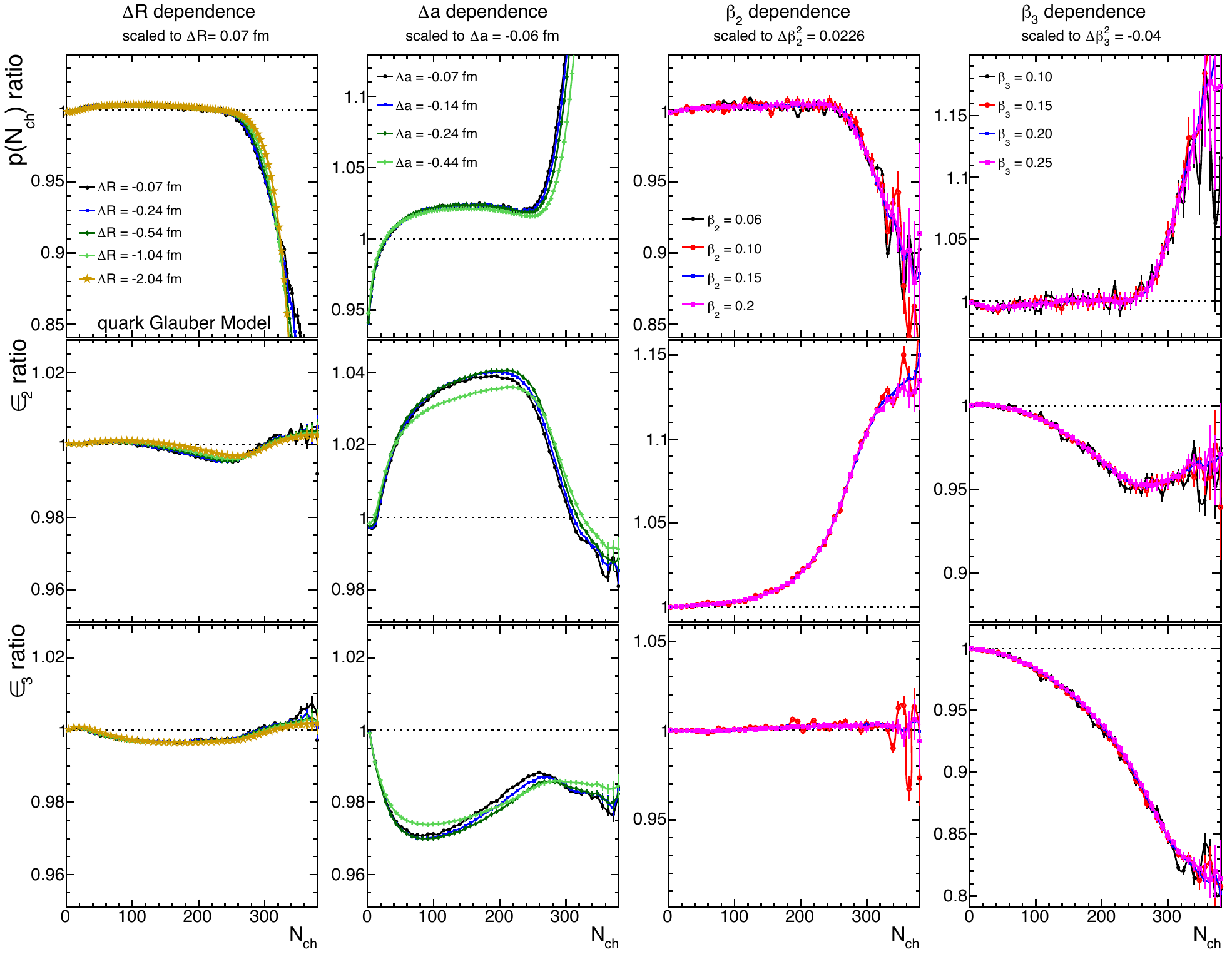}
\caption{\label{fig:5} Information in Fig.~\ref{fig:4} scaled to the default variation in the nuclear structure parameters as indicated in the top of each column. Only selected cases are shown for variations in parameters $R_0$ and $a$ in order to make the plots less busy.}
\end{figure*}

To further understand the behaviors of various response coefficients, we also performed a calculation based on the quark Glauber model discussed in Refs.~\cite{Jia:2021qyu,Jia:2021tzt}, where each nucleon is replaced by three constituent quarks. The distribution of quark-participants $p(\nqp)$ generated for the parameter set of $^{96}$Ru in Table 1 is then convoluted with a negative binomial distribution that describes the production of charged particle for each participant, $\pnbd(n) = \frac{(n+m-1)!}{(m-1)!n!} \frac{\bar{n}^nm^m}{(\bar{n}+m)^{n+m}}$. The convoluted distribution is tuned to match the published $p(\Nch)_{\mathrm{Ru}}$, giving the best fit values of $\bar{n}=0.6535$ and $m=0.7515$. These values are then used to generate the $p(\Nch)$ for all other WS parameters. We then calculate the $\varepsilon_{n}$ as a function of $\Nch$ and obtain the ratios $\varepsilon_{n,\mathrm{Ru}}/\varepsilon_{n,\mathrm{Zr}}$ as an estimator for $v_{n,\mathrm{Ru}}/v_{n,\mathrm{Zr}}$. Figure~\ref{fig:4} shows the ratios of $p(\Nch)$, $\varepsilon_2$ and $\varepsilon_3$ obtained in the quark Glauber model for several values of $R_0$, $a$, $\beta_2$ and $\beta_3$ relative to the default. The corresponding ratios after being scaled to the default variation as listed in Table 1 are shown in Fig.~\ref{fig:5}. Some small deviation from this scaling is observed only when the variations are very large.  

\bibliography{deform}{}
\end{document}